\input harvmac.tex

\def\tilde{\widetilde}

\def\pl#1#2#3{Phys. Lett. {\bf #1B} (#2) #3}

\font\zfont = cmss10 
\font\litfont = cmr6

\def\bigone{\hbox{1\kern -.23em {\rm l}}}
\def\ZZ{\hbox{\zfont Z\kern-.4emZ}}
\def\half{{\litfont {1 \over 2}}}

\def\R{{\bf R}}
\def\Z{{\bf Z}}

\def\S{{\bf S}}
\def\Z{{\bf Z}}
\def\R{{\bf R}}

\Title{hep-th/9603003, RU-96-12, IASSNS-HEP-96/19}
{\vbox{\centerline{Comments on String Dynamics in Six Dimensions}}}
\bigskip
\centerline{N. Seiberg}
\smallskip
\centerline{\it Department of Physics and Astronomy}
\centerline{\it Rutgers University, Piscataway, NJ 08855-0849, USA}
\medskip
\centerline{and}
\smallskip
\centerline{E. Witten}
\medskip
\centerline{\it Institute for Advanced Study}
\centerline{\it Princeton, NJ 08540, USA}
\bigskip
\noindent
We discuss the singularities in the moduli space of string
compactifications to six dimensions with $N=1$ supersymmetry.  Such
singularities arise  from either
 massless particles or  non-critical tensionless
strings.  The points with tensionless strings are sometimes phase
transition points between different phases of the theory.  These
results appear to 
connect all known $N=1$ supersymmetric six-dimensional vacua.

\Date{2/96}

\newsec{Introduction}

In this note we will study some aspects of the dynamics of
supersymmetric compactifications of string theory to six space-time
dimensions.  There are four kinds of supersymmetry algebras in six
dimensions, labeled by the number of supersymmetries of each chirality:

\item{1.} The minimal supersymmetry algebra, $N=1$ or (0,1), is
generated by two spinors of the same chirality.  

\item{2.}  The chiral (0,2) algebra is generated by four spinors of the
same chirality.

\item{3.} The non-chiral (1,1) algebra is generated by two spinors of
one chirality and two spinors of the opposite chirality.

\item{4.}  The maximal supersymmetry algebra, (2,2), is based on four
spinors of one chirality and four spinors of the opposite chirality.  

The simplest object to study is the moduli space of vacua of these
theories.  These are labeled by the expectation values of massless
scalars.  A lot of information about these spaces can be obtained
without a detailed knowledge of the microscopic theory.  The
constraints following {}from supersymmetry and anomaly considerations
turn out to be very powerful.  Together with some information {}from
string theory, these constraints lead to a beautiful and coherent
description of the (2,2), (0,2) and (1,1) compactifications.  In
particular, all known compactifications of string theory (or M-theory)
with these supersymmetries appear to be on the same moduli space of
vacua.

\nref\dmw{M.J. Duff, R. Minasian and E. Witten, hep-th/9601036.}%
\nref\gpo{E.G. Gimon and J. Polchinski, hep-th/9601038.}%
\nref\sen{A. Sen, hep-th/9602010.}%
\nref\dab{A. Dabholkar and J. Park, hep-th/9602030.}%
\nref\strocou{E. Witten, hep-th/9602070.}%
\nref\vafa{C. Vafa, hep-th/9602022.}
\nref\quev{G. Aldazabal, A. Font, L. E. Ibanez, and F. Quevedo, 
hep-th/9602097.}%
\nref\morvafa{D.R. Morrison and C. Vafa, hep-th/9602114.}%
\nref\asgr{P.S. Aspinwall and M. Gross, hep-th/9602118.}%
In this note we will focus on $N=1$ theories, one of the
questions being whether likewise they are all connected.  $N=1$ models
 appear in
compactifications of the heterotic or type I theories on K3 as well as in
other examples.  Such theories were studied recently in
\refs{\dmw-\asgr}.  The massless representations of $N=1$ supersymmetry
with low spins are labeled by their $Spin(4) \sim SU(2)\times SU(2) $
representations
\item{1.} gravity multiplet: $(1,1) + 2(\half,1) + (0,1)$
\item{2.} tensor multiplet: $(1,0) + 2(\half,0) + (0,0)$
\item{3.} vector multiplet: $(\half,\half) +2(0,\half)$
\item{4.} hypermultiplet: $2(\half,0) + 4(0,0)$

There are two kinds of scalars -- those {}from the hypermultiplets and
those {}from the tensor multiplets.  The moduli space of
hypermultiplets is similar to that of the analogous $N=2$
supersymmetric theory in four dimension obtained by compactification
on a two-torus.  It is given by a quaternionic manifold.
Singularities in this space can at least in many cases be associated
with massive vector multiplets becoming massless thus leading to new
massless vectors multiplets and massless hypermultiplets at that
point.  At such singularities the unbroken gauge symmetry is enhanced.
This is the only dynamics of six-dimensional hypermultiplets that can
be seen classically, and therefore the only kind that is possible in
the quantum theory as long as the physics is free in the infrared (and
therefore described at long distances by an effective classical
theory).  Upon reduction to five or fewer dimensions, there are
additional possibilities, since below six dimensions the vector
multiplet contains scalars, and a ``Coulomb branch,'' parametrized by
the expectation values of these scalars, can emanate {}from a
singular point in hypermultiplet moduli space.

There is, however, in six dimensions something that deserves to be
called a Coulomb branch: it is parametrized by the expectation values
of the scalars in the tensor multiplets.  Upon reduction to five or
four dimensions these multiplets become identical to vector
multiplets, so this factor in the moduli space reduces below six
dimensions to an ordinary Coulomb branch.  One of the points of this
note is the identification of the singularities in the six-dimensional
tensor multiplet Coulomb branch with singularities associated to
tensionless strings.  As in four dimensions there can be Higgs
branches emanating {}from these singularities.

In order to explore the Coulomb branch, we should first discuss some
properties of the tensor fields.  In general,
$N=1$ theories in six dimensions have one gravity
multiplet and any number of tensor multiplets, say $n$.  
The gravity multiplet has a
massless state in the (0,1) of $Spin(4)$ while every tensor multiplet
has a massless state in (1,0).  These are two-form gauge fields
$B_{\mu\nu}$ with a one-form gauge invariance $\delta B= d\Lambda$.  The
corresponding
field strengths $H=dB$  are constrained to be self-dual for the
field in the gravity multiplet and anti-self-dual for those in tensor
multiplets.  Except for
 $n=1$, there is no known Lagrangian description of
these theories.

Since the tensor multiplets include gauge fields, the reparametrizations
of the Coulomb branch are quite restricted.  Non-linear transformations
of the tensor fields would not preserve gauge invariance, and
hence their  scalar superpartners
 are also not naturally subjected to arbitrary
nonlinear transformations.  The low energy
theory of the two-forms has  a 
 $O(1,n;{\bf Z})$ duality group, which is a symmetry of the low
energy supergravity\foot{See 
\ref\salam{A. Salam and E. Sezgin, eds., {\it Supergravities In 
Diverse Dimensions} (World Scientific, 1989).}
for review and references.} when accompanied by suitable
transformations of the scalars.  These transformations of the scalars
are the only natural ones to consider.  Note that these are not
necessarily symmetries of the theory but merely the most general
freedom in the parametrization of the field space, preserving the
general form of the low energy supergravity.

As in $N=2$ SUSY in four dimensions, the metric on the hypermultiplet
moduli space is independent of the scalars in the tensor multiplet.
Also, as in four dimensions, the metric on the Coulomb branch and the
kinetic terms for the vector and tensor fields are independent of the
scalars in the hypermultiplets.  They can depend only on the scalars in
tensor multiplets.

Given these rules, there is no supersymmetric coupling that can be
described in classical field theory which could lead to masses for
massless particles by changing the scalars in the tensor multiplets.
This means that such a process cannot occur as long as the physics is
infrared-free.  Conversely, it is impossible for massive particles to
become massless at a particular point on the Coulomb branch, by any
mechanism that can be described at low energies by a free field
theory.  Therefore, singularities in the Coulomb branch are more
subtle.  They necessarily involve the possibility that at the singular
point, the physics is non-free in the infrared, a seemingly rather
exotic possibility since conventional six-dimensional quantum field
theories are believed to be infrared-free.  We will argue that the
singular points on the Coulomb branch are associated with strings that
are becoming tensionless.  These are non-critical string theories;
non-critical here simply means that the graviton is not a mode of the
string, so that the non-trivial infrared physics associated with the
Coulomb branch singularities is occurring in flat six-dimensional
Minkowski space.  For this reason also, these non-trivial infrared
fixed points very possibly may obey the general axioms of quantum
field theory, for instance possessing a local energy-momentum tensor
and other local fields, though apparently  not being realized as
Lagrangian field theories in any natural way.  (This last point has
been stressed to us by L. Susskind.)

It is helpful to have a concrete example in mind.  Consider the simple
case with $n=1$, where it is easy to describe the system in terms of a
Lagrangian.  The free Lagrangian in the Einstein frame is
\eqn\finvlag{-\half e^{-2\phi} (dB- \omega_m)\cdot (dB -\omega_m) - B
\wedge d\omega_e}  
where $e^\phi$ is the coupling constant.  Here $\omega_m$ and
$\omega_e$ are linear combinations of gauge and gravitational
Chern-Simons three-forms, and the four-forms $J_e=-d\omega_e$ and
$J_m= -d\omega_m$ can be interpreted as electric and magnetic
sources\foot{A similar Lagrangian in four dimensions with $B$ a one
form describes the coupling of Maxwell fields to electric and magnetic
sources.}.  This interpretation becomes obvious {}from the equation of
motion and the Bianchi identities for $H=dB-\omega_m$, which read
\eqn\soureq{\eqalign{&d^*e^{-2\phi} H= J_e \cr
&dH= J_m . \cr}} 
The charged objects are strings whose electric and magnetic charges are
defined as 
\eqn\charde{\eqalign{&n_e=\int_\Sigma J_e\cr
&n_m=\int_\Sigma J_m\cr}}
where $\Sigma$ is a four manifold transverse to the string.  Note that
for non-zero charges, $\omega_e$ or $\omega_m$ should decay sufficiently
slowly at infinity.  In this case the duality group is $O(1,1;{\bf Z})$.
The interesting duality transformation expresses the Lagrangian
\finvlag\ in terms of the dual field strength
$\tilde H={}^*e^{-2\phi}H$.  This duality transformation exchanges the
two equations in \soureq\ and inverts the coupling constant.  Since it
reverses the sign of the anti-self-dual tensor, by supersymmetry it
should also reverse the sign of the scalar in that multiplet.
Therefore, we can identify $\phi$ in \finvlag\ as that scalar.

The supersymmetry algebra, because of its chiral nature,
 can be extended by a central
charge that transforms as a Lorentz-vector but does not admit
Lorentz-scalar central charges.  Of course, a Lorentz-vector
central charge would be carried by a string, not a particle.
This is in keeping with the fact that the supergravity multiplet
contains a two-form, and not an ordinary gauge field; the two-form
naturally couples to a string, not a particle.  
A standard BPS argument leads to an inequality for the tension of any
string that couples electrically or magnetically to the $B$ field:
\eqn\tenine{T \ge |n_e a_e + n_m a_m|.}
Here $n_e$ and $n_m$ are the electric and magnetic charges of the
string and $a_e$ and $a_m$ depend on the coordinates on the moduli
space.  As in the analogous four-dimensional case, $a_e$ and $a_m$
 can depend only on the scalars in the tensor multiplets and not on the
scalars in the hypermultiplets.  In the example \finvlag\ above
$a_e=e^{\phi}$ and $a_m=e^{-\phi}$.  Such BPS saturated strings were
studied in
\ref\strs{M.J. Duff, S. Ferrara, R.R. Khuri and J. Rahmfeld,
hep-th/9506057.}. 

As we move on the moduli space the tensions of the strings can vary.
A string labeled by $(n_e,n_m)$ can become tensionless at the point
where $n_e a_e + n_m a_m=0$ and lead to a singularity in the moduli
space.  The situation is very similar to the analogous one in $N=2$ in
four dimensions.  There masses of particles (rather than tensions of
strings) are controlled by a BPS formula on the Coulomb branch.  A
singularity in the Coulomb branch is associated with massless
particles.  At the vicinity of this singular point we can focus on the
string which becomes tensionless and on the unique tensor multiplet
which couples to it.  The other modes of the theory are either much
heavier or decouple {}from that string.  Then an effective Lagrangian
like \finvlag\ gives a good description of the massless modes in the
vicinity of the singularity.  The singularity is at $\phi=\phi_0$ with
$e^{-2\phi_0}= -{n_e \over n_m }$.  Furthermore, only the
anti-self-dual field $H - e^{-2(\phi-\phi_0)} {}^*H$ couples to the
non-critical string which becomes tensionless there while the
self-dual field $H + e^{-2(\phi-\phi_0)}{}^*H $ decouples.

To make this discussion of the vicinity of the singular point $\phi_0$
more precise, we should define an appropriate scaling limit in which
one sees the non-trivial infrared fixed point.  In the
previous discussion we have set the Planck scale to one, but now we will
be more explicit and will denote it by $M_p$.  Then, the appropriate
scaling limit is $\phi \rightarrow \phi_0$ combined with $M_p
\rightarrow \infty$ holding $T=M_p^2 (\phi-\phi_0)$ fixed.  In this 
limit all the massive modes, whose mass is of order $M_p$, and most of
the massless modes decouple.  (If this transition occurs at weak
string coupling, there are also modes whose mass is much smaller than
$M_p$.  They decouple because of the small string coupling.)  We are
left with the modes of the non-critical string which becomes
tensionless at the singularity (its tension $T$ is finite in the
scaling limit) and one tensor multiplet whose scalar $\phi-\phi_0$
vanishes at the critical point.  There might also be additional
``relevant'' massless fields in some cases.  This scaling theory has
an accidental duality symmetry (which might be a symmetry of the full
theory) under which $H\leftrightarrow e^{-2(\phi-\phi_0)}{}^*H $ and
$(\phi-\phi_0)\leftrightarrow -(\phi-\phi_0)$.  It is crucial that the
gravity multiplet also decouples in this limit.  This follows {}from
the fact that in the limit $M_p \rightarrow \infty$ gravity decouples
{}from all finite mass (or finite tension) objects.  Alternatively, it
follows {}from the fact that the self dual field $H +
e^{-2(\phi-\phi_0)}{}^*H $ which is in the gravity multiplet
decouples.

As in $N=2$ in four dimensions, one can conceive of the possibility of
Higgs branches emanating {}from the singular point.  On the Higgs
branch some tensor multiplets are ``Higgsed.''  Even without knowing
the details of the microscopic theory, anomaly considerations
constrain the transition {}from the Coulomb branch to the Higgs
branch.  Denoting the number of tensor multiplets by $n$, the number
of hypermultiplets by $H$ and the number of vector multiplets by $V$,
anomaly cancelation implies
\eqn\anomac{29n+H-V=273.}
Therefore, if $V$ is unchanged and exactly one tensor is Higgsed, there
should be 29 more hypermultiplets on the Higgs branch than on the
Coulomb branch.  Other transitions where the value of $V$ also changes
are also conceivable.

In section two we will discuss compactifications of the $SO(32)$ type
I and heterotic strings to six dimensions while in section three we
will discuss the $E_8\times E_8$ heterotic string.  In section four we
will examine compactifications of M theory on $K3 \times (\S^1/\Z_2)$
which do not correspond to perturbative vacua of the heterotic string.
Using these compactifications we will interpolate between different
vacua of the heterotic string.  In section five we will discuss the
strong coupling transition and will mention some vacua of string
theory which do not have a dilaton.

\newsec{Compactifications of the $SO(32)$ theory on K3}

When the type I theory is compactified on a K3 surface the low energy
theory has $n=1$ and the Lagrangian is of the form \finvlag.  For Type
I, the scalar $\phi$ in the tensor multiplet is essentially the volume
of K3 measured in the type I metric
\eqn\tensorI{v_I= e^{-2\phi}.}
In particular $\phi$ is independent of the ten dimensional dilaton $D_I$.

The hypermultiplets include fields which are charged under the gauge
symmetries and also some moduli of K3.  In particular, one combination
of $D_I$ and $\phi$ transforms in a hypermultiplet.
Straightforward dimensional reduction shows that the appropriate
combination is $e^{D_I}/v_I^{1/2}$, the effective six-dimensional
string coupling constant.

The coupling of the two-form $B$
\eqn\bterms{-{1 \over 2} e^{-2\phi} (dB-\omega_3)\cdot (dB-\omega_3) 
- B \wedge d\tilde \omega_3  }
is correlated with the gauge kinetic terms
\ref\sagnotti{A. Sagnotti, \pl{294}{1992}{196}, hep-th/9210127.}
\eqn\invlag{- \sum_i (v_i e^{-\phi} + \tilde
v_i e^{\phi}) \tr F_i \cdot F_i}
through
\eqn\defom{\eqalign{&d\omega=-{1 \over 16 \pi^2} \sum_i v_i 
\tr F_i \wedge F_i \cr
&d \tilde \omega=-{1 \over 16 \pi^2} \sum_i \tilde v_i 
\tr F_i \wedge F_i}}
and $i$ labels different gauge groups.  We recognize the two linear
combinations of Chern-Simons three-forms, $\omega$ and $\tilde \omega$
as $\omega_m$ and $\omega_e$ in \finvlag.  Instantons in the gauge
fields are localized in four dimensions and hence they are strings in
six dimensions.  We see that an instanton in the gauge group labeled
by $i$ with instanton number $n_i$ leads to a string with charges
$(n_e=\tilde v_in_i, n_m= v_in_i)$.  Since the instanton field
strength is self-dual in four dimensions, it preserves half the
supersymmetries and hence this string is BPS saturated.  The tension
of this string is, {}from the central charge formula, $T= n_i(v_i
e^{-\phi} + \tilde v_i e^{\phi})$.  Note that this value of the
tension can also be deduced {}from the value of the action of this
configuration; it comes {}from the coefficient of the gauge kinetic
term.  If the charges are such that the tension can vanish, at some
point, the gauge coupling diverges at that point.  This is the
transition point discussed in \dmw.  Here we identify it as associated
with tensionless non-critical strings.

To some extent, the fact that the gauge coupling goes to infinity
enables one to  get an intuitive picture of how infrared non-trivial
physics comes about. In six-dimensional Yang-Mills theory, the bare
gauge coupling $g$ has dimensions of length.  The theory is believed
to be infrared-free, meaning that it is weakly coupled at distances
much bigger than $g$.  As $g\to\infty$, the length scale above
which weak coupling prevails becomes greater and greater, and at
the critical point of the tensor moduli space, $g=\infty$ and there
is no such length scale: the physics is non-trivial in the infrared.

Consider for example the special point in the moduli space with gauge
symmetry $SO(32) \times Sp(24)$.  For $Sp(24)$ we have $v_{24}=0$ and
$\tilde v_{24}= 2$, and for $SO(32)$ $v_{32}=1$ and $\tilde
v_{32}=-2$.  Since $v_{24}=0$, an $Sp(24) $ instanton is purely
electric $(n_e=2,n_m=0)$.  {}From a ten dimensional point of view, the
$Sp(24)$ gauge fields exist on a five-brane which fills the
non-compact dimensions.  Instantons on such branes were studied in
\ref\douglas{M. Douglas, hep-th/9602098.},
where it was shown that when they become small they can leave the brane
as ``elementary'' strings.  This is consistent with the fact that they
are purely electric and therefore can be identified with the dual
heterotic string.  An $SO(32)$ instanton has $(n_e=-4,n_m=2)$ (an
overall factor of 2 appeared because of the index of the vector
representation of $SO(32)$).  In ten dimensions the small $SO(32) $
instanton is a Dirichlet five-brane.  
The six-dimensional instanton is obtained
by wrapping this five-brane on K3 to yield a string in six dimensions.
We see that for $e^{2\phi_0}= 2$ this string becomes tensionless and
leads to a singularity.  Note that this is not a strong coupling
singularity.  It happens at a point where the sigma model is strongly
coupled (the sigma model coupling $v_I^{-1/4}$ is of order one there)
but the string coupling constant $e^{D_I} $ can be arbitrarily small.

An interesting subtlety has to be mentioned here.  These two strings have
several collective coordinates.  One of them is associated with the
width of the string -- the size of the instanton.  This collective
coordinate is non-compact and its quantization leads to a continuum.
Therefore, there is no gap in the spectrum of the string.  It might be
that this continuum of states can be interpreted as a brane in its
ground state surrounded by soft non-Abelian gauge bosons.  However, as
explained in \douglas, the world-sheet theory of the $Sp(24)$ instanton
has another branch where there is clearly a gap in the spectrum.  It is
in this branch that the string can be identified as the elementary
heterotic string.  The other strings may also have such alternative
branches.

Under heterotic - Type I duality, the Type I model just discussed
can be described as an $SO(32)$ heterotic string, for which
the singularity discussed 
above occurs at strong coupling.  The scalar in the tensor multiplet is
given in the heterotic string description by 
\eqn\tensorI{v_he^{-2D_h} = e^{-2\phi}}
where $v_h$ is the volume of K3 in the heterotic string units and
$D_h$ is the ten dimensional heterotic dilaton.  The various strings
discussed in the type I compactification are easily identified in the
heterotic theory.  The purely electric string is the fundamental
heterotic string while the dyonic string is a solitonic five-brane
wrapped over the K3.

\newsec{Compactifications of the $E_8\times E_8$ heterotic string on K3}

Compactifications of the $E_8\times E_8$ heterotic string on K3 also
give $N=1$ models in six dimensions.  These models are associated with
$(0,4)$ superconformal field theories.  They are labeled, at least
initially, by the instanton numbers $n_1$ and $n_2$ in the two $E_8$
factors.  In this context the total instanton number
\eqn\instco{n_1 +n_2 =24.}
Classically, smooth $E_8$ instantons with $n=1,2,$ or 3 do not exist.
Therefore, we have the following possibilities:

\item{1.}  $n_1=24$, $n_2=0$.  At the generic point in the moduli 
space the instantons completely break the first $E_8$ and the unbroken
gauge group is the second $E_8$.  At a special point (the standard
embedding) the first $E_8$ is broken to $E_7$.  At that point the
theory has 10 hypermultiplets in the $({\bf 56,1})$ of $E_7\times E_8$
as well as 65 gauge invariant hypermultiplets and a single
tensor multiplet ($n=1$) which includes the dilaton.  It is easy to
see that equation \anomac\ is satisfied and the anomaly polynomial
factorizes as it should
\nref\erler{J. Erler, J. Math. Phys. {\bf 35} (1994) 1819, 
hep-th/9511030.}%
\nref\schwarz{J.H. Schwarz,  hep-th/9512053.}%
\refs{\erler,\schwarz}
\eqn\anomalyo{(\tr R^2 - A_1 -A_2)(\tr R^2 - 6 A_1 +  6A_2) }
with $A_1=\tr F_1^2/6$ for the $E_7$ factor and $A_2=\tr F_2^2/30$ for
the $E_8$ factor.  At more generic points on the moduli space 
$E_7$ can be Higgsed to a subgroup.

\item{2.}  $n_1,n_2 \ge 4$ with $n_1+n_2=24$.  The instantons break 
the two $E_8$ factors.  At special points, where all the instantons in
each $E_8$ are at an $SU(2)$ subgroup, the unbroken gauge group is
$E_7\times E_7$.  There are also $n_1-4$ half hypermultiplets in
$({\bf 56,1})$, $n_2-4$ half hypermultiplets in $ ({\bf 1,56})$ as
well 62 gauge invariant hypermultiplet and a tensor multiplet.  Again,
equation \anomac\ is satisfied and the anomaly polynomial factorizes
\eqn\anomalyt{(\tr R^2 - A_1 -A_2)(\tr R^2 - {n_1-12 \over 2} A_1 
+ {n_1-12 \over 2}A_2) } 
with $A_i=\tr F_i^2/6$ for the two $E_7$'s.  At more generic points
the two $E_7$ groups can be Higgsed.  To break $E_7$ completely we
need at least 10 instantons.  Therefore in the theories with
$(n_1,n_2)=(14,10),~(13,11)$ and $(12,12)$ the gauge symmetry can be
completely broken.

As in \invlag, the anomaly polynomial \anomalyt\ allows us to find the
gauge kinetic terms.  They are proportional to
\eqn\invlago{- (e^{-\phi} +{n_1-12 \over 2} e^{\phi}) \tr F_1 \cdot F_1 
- (e^{-\phi} +{n_2-12 \over 2} e^{\phi}) \tr F_2 \cdot F_2}
where $F_1$ and $F_2$ are the field strengths of subgroups of the the
first and the second $E_8$ factors.  Assuming, without loss of
generality, that $n_1\ge n_2$, the gauge coupling of the first
$E_8$ is always finite while the other gauge coupling diverges at
$\phi=\phi_0$ with
\eqn\singloc{e^{-2\phi_0} = {n_1-12 \over 2}.}
At that point the string associated with small instantons in that
gauge group becomes tensionless.  We will return to this singularity
below.

It can actually be shown using $T$-duality that the $(16,8)$ model
is equivalent to the $SO(32)$ heterotic string with standard embedding
of the spin connection in the gauge group \ref\ewitten{E. Witten,
to appear.}, and similarly the $(12,12)$ model is equivalent to 
that discussed in \gpo.  It also seems likely, given results in
\morvafa, that the $(14,10)$ and $(12,12)$ models are related purely
at the level of conformal field theory, by some sort of $T$-duality.

\newsec{Compactifications of M-theory on $K3 \times (\S^1/\Z_2)$}

The eleven dimensional description (M-theory) naturally leads to
additional $N=1$ supersymmetric vacua which do not have a perturbative
heterotic string theory interpretations.  Some of them were discussed
in \sen\ and have a perturbative type II interpretation \dab.  Others,
which we will focus on below, were mentioned in \dmw.  The common fact
about these vacua is that they have more than one tensor multiplet,
$n>1$.

Consider a compactification of M-theory on 
$K3 \times ({\bf S}^1/{\bf Z}_2)$.  The
gauge fields of the two $E_8$ factors are on two different ``end of
the world'' nine-branes
\ref\horwi{P. Horava and E. Witten, hep-th/9510209.}.
We can now embed $n_1$ instantons in one $E_8$ factor and $n_2$
instantons in the other.  However, there is an extra
possibility not seen in the perturbative heterotic
compactifications:  we can also add $n-1$ five-branes which are located
at points on ${\rm K3}\times {\bf S}^1/{\bf Z}_2$, and fill the
non-compact six dimensions.  The location of such a five-brane
on 
${\rm K3}\times {\bf S}^1/{\bf Z}_2$  
is  labeled by five real parameters.  These parameters form
a hypermultiplet (the coordinate on K3) and  a tensor multiplet (the
coordinate in $\S^1/\Z_2$).  Therefore, together with the tensor
multiplet which includes the dilaton (the distance between the two
nine branes), such vacua have $n$ tensor multiplets.  As these 
five-branes are sources for the antisymmetric gauge fields, the equation
$n_1+n_2=24$ is now replaced by
\eqn\newcon{n_1+n_2+n-1=24.}
In other words, an instanton {}from one of the nine-branes can be
replaced by a five-brane \dmw.

We can easily extend the anomaly considerations to this more general
case (see also the discussion in
\ref\strong{E. Witten, hep-th/9602070.}):

\item{1.} $n_1=n_2=0$.  Here the unbroken group is $E_8\times E_8$ and 
there are $n=25$ tensor multiplets (24 {}from the branes and one which
includes the dilaton) and 44 hypermultiplets (24 {}from the locations of
the branes and 20 moduli of the K3 metric).  This spectrum satisfies
\anomac.

\item{2.}  $n_2=0$, $n_1 \ge 4$.  At special points in the moduli space 
the instantons break the gauge group to $E_7 \times E_8$.  There are
also $n=25-n_1$ tensor multiplets and $n_1-4$ half hypermultiplets in
$ ({\bf 56,1})$ and $n_1 +41$ gauge invariant hypermultiplets
($24-n_1$ {}from the locations of the five-branes, $2n_1-3$ {}from the
moduli of the gauge bundle and 20 {}from the moduli of the metric).
Again, \anomac\ is satisfied.  At more generic points on the moduli
space the $E_7 \times E_8$ symmetry can be Higgsed.

\item{3.}  $n_1,n_2 \ge 4$.  At special points the gauge symmetry is 
$E_7 \times E_7$ with $n=25-n_1-n_2$ tensors, $n_1-4$ half
hypermultiplets in $ ({\bf 56,1})$, $n_2-4$ half hypermultiplets in $
({\bf 1, 56})$ and $n_1+n_2 +38$ hypermultiplets ($24-n_1-n_2$ {}from
the locations of the 5 branes, $2n_1+2n_2-6$ deformations of the two
gauge bundles and 20 metric moduli).  As before, \anomac\ is satisfied
and at more generic points the gauge symmetry can be Higgsed down.

Equation \anomac\ is a necessary condition for cancelling the anomaly.
Other conditions come {}from examining the anomaly eight form.  In all
of these cases it is 
\eqn\anomalytt{(\half \tr R^2 - A_1)({n_1-8 \over 4}\tr R^2 -{n_1-12
\over 2} A_1) +  (\half \tr R^2 - A_2)({n_2-8 \over 4}\tr R^2 -{n_2-12
\over 2} A_2) }
with $A_i=\tr F_i^2/6$ for an unbroken $E_7$ and $A_i=\tr F_i^2/30$
for an unbroken $E_8$ ($i=1$ for subgroups of the first $E_8$ factor
and $i=2$ for the second).  In the special cases with $n=1$ it
coincides with \anomalyt.  The form of \anomalytt\ shows that it is a
sum of terms associated with the two different boundaries.  Note that
for $n>1$ the anomaly does not have to factorize because there are
more two-forms to cancel it \sagnotti.

These new vacua of string theory which are not perturbative heterotic
vacua allow phase transitions between different values of $(n_1,n_2)$.
The vacua with given initial values of $(n_1,n_2)$ 
can be thought of as the Higgs branch.  As an
$E_8$ instanton shrinks, a singularity is found.  That singularity
can be interpreted as resulting {}from a five-brane stuck to the end
of the world nine-brane.  It is very plausible that another phase
\nref\haga{O.J. Ganor and A. Hanany, hep-th/9602120.}
 is obtained when the
five-brane moves away {}from the boundary.  
Arguments for this have been given in \haga.
Since we gain a tensor
multiplet when the  five-brane leaves the boundary, the new 
phase can be interpreted as a Coulomb branch.  This phase
transition is, therefore, rather like what was discussed
in general terms in the introduction.  On the Coulomb
branch there is a BPS saturated string which is an eleven-dimensional
two-brane which stretches {}from the five-brane to the boundary (we
know {}from \horwi\ that two-branes can end on the boundary and {}from 
\ref\andy{A. Strominger, hep-th/9512059.}
that they can end on five-branes).  At the singularity this string
becomes tensionless.  After being emitted to the bulk, the five-brane
could possibly travel to the other end of ${\bf S}^1/{\bf Z}_2$
and be absorbed on the other boundary.
So  in M-theory, assuming such transitions are really
possible, all values of $(n_1,n_2) $ are connected.

Other singularities in the moduli space occur when the number
of tensors is  $n\ge 3$; i.e.\ when
there are at least two five-branes in the bulk.  These 
singularities correspond to two
five-branes approaching each other.  Again, this is a singularity in the
Coulomb branch which is associated with non-critical strings.  This
non-critical string was first discussed in
\ref\wittenstring{E. Witten, hep-th/9507121.}
where it appeared near a singularity of the IIB theory compactified on
K3 yielding a (0,2) supersymmetric theory in six dimensions.  This
singularity was interpreted as arising {}from two five-branes
approaching each other in
\nref\wittenbranes{E. Witten, hep-th/9512219.} \refs{\andy,\wittenbranes}.
In \sen, this singularity appeared in a theory with only (0,1)
supersymmetry like the theories discussed here.

The scaling theory associated with this non-critical string, which
we will call the non-critical Type II string,  is
different {}from that of the small $E_8$ instanton.  One way to see
that is to note that this non-critical string has twice as many
supersymmetries as the other -- the low energy scaling theory around
the singularity has accidental supersymmetry.    
It has $(4,4)$ world-sheet supersymmetry (as is clear from its origin
in Type II) in contrast to the $(0,4)$ world-sheet supersymmetry
of the string related to small $E_8$ instantons.
Another difference between
the two cases
 is that the singularity associated with the non-critical Type II
string is of real codimension five while the
other is of real codimension one.  One needs to tune a hypermultiplet
(the distance between the two five-branes in K3)
and the scalar in a tensor multiplet (the distance between the two
five-branes in ${\bf S}^1/{\bf Z}_2$)
to find the non-critical Type II string.
This is consistent with the (0,2) space-time
supersymmetry of this string; one needs to tune a tensor multiplet of
(0,2) supersymmetry, which includes five real scalars, to find the
string; the five scalars form in $(0,1)$ space-time supersymmetry
a hypermultiplet and part of a tensor multiplet.
This might seem in contradiction with the fact mentioned in
the introduction that the BPS bound is independent of hypermultiplets.
Why is it then that we have to tune a hypermultiplet to find the
tensionless string?  The answer is that this string is BPS-saturated
in (0,2) supersymmetry but it is not BPS-saturated in (0,1)
supersymmetry. 
Therefore, when the scalar in the tensor multiplet is
tuned to a point where the BPS bound vanishes but the hypermultiplet
is at a generic point, this string has positive tension.  However, by
tuning also the hypermultiplet it can become tensionless.

One can show that, unlike the case of the small $E_8$ instanton, there
is no Higgs branch emanating {}from this singularity.  One way to see
that is to consider the type IIB string compactified on K3 where the
same non-critical string appears.  In this case anomaly considerations
fix the number of (0,2) tensor multiplets and there is no branch where
this number is reduced.

The singularity in the moduli space associated with this transition
is the orbifold singularity  $\R^5/\Z_2$.  The identification
by $\Z_2$ reflects Bose symmetry of the two five-branes.  Therefore,
when one adjusts the parameters to see make the tension of this string
vanish, one sees an enhanced $\Z_2$ symmetry of the physics.  
We return to 
this point in the next section.

\newsec{First Look At The Strong Coupling  Singularities}

We have mentioned above other kinds of singularities in the Coulomb
branch -- those associated with the diverging gauge coupling.  In
this section we are going to take a first look at them.  For
simplicity we limit the presentation to the case of one tensor
multiplet and to
compactifications which have a perturbative heterotic string
description.  They are labeled by the two instanton numbers
$(n_1,n_2)$, and we will assume without loss of generality that
$n_1 \ge n_2$.  Then the gauge coupling of a subgroup of the second
$E_8$ diverges at the value of $\phi$ satisfying \singloc.  At that
point instantons in that group lead to tensionless strings and
complicated dynamics can arise.  Independent of what this dynamics is,
it is clear that the gauge bosons in the first $E_8$ are spectators
which do not participate in it. This is obvious, for instance, in the
M-theory description, where the two $E_8$'s are supported on K3's that
can be arbitrarily far apart (while making one of them larger), only
one of which is affected by the singularity.

What kind of string can arise at the strong coupling singularity?
For $n_2\leq 9$, there is a generic unbroken gauge group $H$ , whose
coupling goes to infinity at $\phi=\phi_0$.  Therefore, in this
case, there must be a singularity at $\phi=\phi_0$ for generic
values of the other fields.  
The fact that the string becomes tensionless upon adjusting $\phi$
and nothing else strongly suggests that it is BPS-saturated,
with a tension controlled purely by $\phi$.  The instanton in the
strongly coupled gauge group will indeed do the job if there is nothing
else.  For $n_1=n_2=12$, there is no value
of $\phi$ at which gauge couplings diverge \dmw, so one has no reason
to expect a singularity to arise on adjusting $\phi$ for  generic
values of the other fields; the existence of such a singularity
would really  contradict the duality of \dmw.  For $n_2=11,10$,
such general arguments do not make it clear what happens at $\phi=\phi_0$.

If a singularity does arise at $\phi=\phi_0$, can one continue beyond
it in $\phi$ space?  The obvious intuitive idea, analogous to
$R\to 1/R$ symmetry in conformal field theory, is that the physics
at $\phi>\phi_0$ might be isomorphic to that at $\phi<\phi_0$, related
by a $\Z_2$ symmetry.  This is, however, possible only if
$\phi_0$ is the self-dual value of the dilaton, that is if $\phi_0=0$,
since the only transformation of the dilaton that leaves invariant
the low energy supergravity is $\phi \to -\phi$.  A look at \singloc\
shows that $\phi_0=0$ if and only if $n_2=10,$ $n_1=14$.  Thus,
only at this value of the instanton number is the naive idea of
finding isomorphic physics beyond the strong coupling singularity
conceivable.

Note further that it has been argued {}from F-theory \morvafa\ that
the $n_2=10$ and $n_2=12$ models are actually the same (equivalent on
the nose, not just connected by a series of phase transitions).  The
$n_2=12$ theory does indeed have a $\phi\to -\phi$ symmetry, so if
they are equivalent, the $n_2=10$ theory must have one also.  Thus,
the continuation beyond the singularity must hold in this case.  Since
the $n_2=12$ theory does not have a strong coupling singularity at
$\phi=0$ for generic values of the hypermultiplets, the same must be
true at $n_2=10$; as we noted two paragraphs ago, there is no
contradiction here, since the $n_2=10$ theory has generically no
unbroken gauge group.  On the other hand, since upon adjusting some
additional parameters the $n_2=10$ theory does have unbroken gauge
symmetries whose couplings diverge at $\phi=0$, it must be that it
develops a tensionless non-critical string upon adjusting $\phi$
together with some other parameters.  Since more parameters than
$\phi$ are involved, this string must not be BPS-saturated at generic
values of the parameters, so it must carry a $(4,4)$ and not just
$(0,4)$ world-sheet supersymmetry.  In fact, {}from F-theory one can
see \ewitten\ that the relevant string at the strong coupling point of
the $n_2=10$ model is the non-critical
Type II string of \wittenstring, which also
entered above in discussing models with more than one tensor
multiplet; as we discussed above, to make this string tensionless, one
must adjust one hypermultiplet as well as one tensor
multiplet.\foot{The key point in getting the non-critical
Type II string here
{}from F-theory is that the F-theory description of this model
involves \morvafa\ compactification on the Hirzebruch surface ${\bf
F}_2$, and the singularity arises when an exceptional two-sphere of
self-intersection number $-2$ collapses; the resulting singularity
looks just like Type IIB theory at an ${\bf A}_1$ singularity, which
gives the Type II string as in \wittenstring.}  We explained above
that no Higgs branch emanates {}from a singularity of this nature.
As remarked at the end of the last section, a $\Z_2$ symmetry
appears at a point at which the non-critical Type II string becomes
tensionless.  In the case under discussion, this $\Z_2$ is
the   strong-weak coupling $\phi\to -\phi$ symmetry of the $(14,10)$ 
model.

For further insight, we look at the structure of the gauge kinetic
energy for $n_2=10$, as was first done in \quev.  Using \invlago, the
gauge kinetic terms are
\eqn\invlagoa{- (e^{-\phi} + e^{\phi}) \tr F_1 \cdot F_1
- (e^{-\phi} - e^{\phi}) \tr F_2 \cdot F_2.}  
The $F_1$ fields are expected to be spectators in the strong
coupling transition, for reasons given above.  The fact
that their coupling is invariant under $\phi\to -\phi$ 
is, as noted in \quev,
 compatible with the existence of a $\phi\to -\phi$ symmetry
under which they are spectators.
The gauge fields $F_2$ have a coupling that diverges at the self-dual
point and are definitely {\it not} spectators.  
(The $F_2$ gauge fields are only massless if some parameters
in addition to $\phi$ are adjusted, to restore a gauge symmetry
that generically is spontaneously broken.  To make sense of the
singularity in the $F_2$ gauge coupling, the  parameters in question
must include the extra hypermultiplet that must be adjusted to make
the non-critical Type II string tensionless; this can be verified using
F-theory.)  
It must be that the formula \invlagoa, which holds in the region 
containing
the perturbative heterotic string of $n_2=10$, should be modified
so that the coefficient of $\tr F_2\cdot F_2$ is really
$\left|e^{-\phi}-e^{\phi}\right|$.  We do not understand very
well how this comes about, though since the gauge fields involved
are not spectators there does not appear to be a contradiction.

The fact that the $F_2$ gauge coupling diverges
at $\phi=0$ is related to the fact that there is no $\tr R^2 \tr F_2^2$
term in the anomaly eight-form; this in turn means that the $F_2$
gauge fields couple to hypermultiplets with the same quadratic
Casimir operator as that of the vector multiplet.  But, upon
reduction to four dimensions, that is the condition for vanishing
beta function!  
This seems like a rather interesting way to generate a large
class of finite $N=2$ models from string theory.
The $\phi\to -\phi$ symmetry
of $n_2=10$, and the behavior near the strong coupling
singularity, are thus very plausibly related, upon toroidal
compactification  to four dimensions, to interesting behavior
of finite $N=2$ models.

Could there be for any values of $n_2$ a Higgs branch emanating {}from
the strong coupling singularity?  Since the number of tensor
multiplets is $n=1$ for perturbative heterotic strings, a Higgs branch
would have $n=0$, no tensor multiplets at all.  It would be a branch
without a dilaton, something one would like to find in four
dimensions!  Since in such a branch, there is only the self-dual
two-form in the supergravity multiplet, anomaly cancelation requires
\sagnotti\ that the anomaly eight-form should not only factorize 
but should be a perfect square.  Since the gauge fields in the first
$E_8$ are spectators in such a transition, they should survive in the
Higgs branch.  So we can test for the occurrence of such a branch by
looking at the anomaly form including the Riemann tensor and $F_1$
(but not the $F_2$ fields, which might get masses on the Higgs
branch).  It is easy to find, using \anomac, that the relevant anomaly
is
\eqn\anomalyttt{{9\over 8} (\tr R^2)^2  - {n_1-10 \over 2}\tr 
R^2 A_1 +{n_1-12 \over 2} A_1^2.}
This is a perfect square only for $n_1=16, 13$, and thus
$n_2=8,11$.  For other values of $n_2$ there can be no Higgs branch.

We do not know the interpretation of this result for $n_2=8$, but for
$n_2=11$ one can show using F-theory that the Higgs branch does indeed
exist \ewitten.  In fact, the Higgs branch in this case is simply
F-theory on ${\bf P}^2$ (which has no tensor multiplets as is clear
{}from \vafa).  The transition to the Higgs branch is made by blowing
down the exceptional curve $C$ in the Hirzebruch surface ${\bf F}_1$
(the right surface for $n_2=11$ as explained in \morvafa) to go to
${\bf P}^2$.  The non-critical string is made by wrapping a Type IIB
three-brane around $C$, and carries a rank eight current algebra,
strongly suggesting that it coincides with the string seen in M-theory
when a five-brane approaches the boundary; that string also carries
a rank eight current algebra, and seems to
have a Higgs branch (small $E_8$ instanton), as discussed in \haga\
and above.  Actually, the M-theory transition where a five-brane
enters or leaves the boundary can be seen in F-theory by blowing up
and blowing down points; by a sequence of such blow-ups and
blow-downs, to give transitions {}from ${\bf F}_n$ to ${\bf F}_{n\pm
1}$, one can see the M-theory process in which an instanton is
emitted at one boundary as a five-brane and then reabsorbed at the
other boundary, changing $n_1$ and $n_2$.

\vskip36pt
\centerline{{\bf Acknowledgements}}
 
We would like to thank J. Polchinski, S. Shenker, and L. Susskind for
discussions.  This work was supported in part by DOE grant
\#DE-FG05-90ER40559 and in part by NSF grant \#PHY95-13835.
 
\listrefs
 
\end